\documentclass[aps,prl,reprint,nopacs,amsmath,amssymb,superscriptaddress]{revtex4-1}%

\usepackage{multirow}
\usepackage{bigstrut}
\usepackage{amssymb}
\usepackage{amsmath}
\usepackage{graphicx}
\usepackage{dcolumn}
\usepackage{bm}
\usepackage{CJK}
\usepackage{relsize}
\usepackage{cancel}
\usepackage{color}
\usepackage{setspace}
\usepackage{mathptmx}

\begin{document}

\title{Impurity-assisted tunneling magnetoresistance  under  weak magnetic field}

\author{Oihana Txoperena}
\altaffiliation{These authors contributed equally to this work.}
\affiliation{CIC nanoGUNE, 20018 Donostia-San Sebastian, Basque Country, Spain}
\author{Yang Song}
\altaffiliation{These authors contributed equally to this work.}
\affiliation{Department of Electrical and Computer Engineering, University of Rochester, Rochester, New York, 14627, USA}
\author{Lan Qing}\affiliation{Department of Physics and Astronomy, University of Rochester, Rochester, New York, 14627, USA}
\author{Marco Gobbi}\affiliation{CIC nanoGUNE, 20018 Donostia-San Sebastian, Basque Country, Spain}\affiliation{Universit\'e de Strasbourg, Institut de Science et d'Ing\'enierie Supramol\'eculaires (I.S.I.S.), 67083 Strasbourg, France}
\author{Luis E. Hueso}\affiliation{CIC nanoGUNE, 20018 Donostia-San Sebastian, Basque Country, Spain}\affiliation{IKERBASQUE, Basque Foundation of Science, 48011 Bilbao, Basque Country, Spain}
\author{Hanan Dery}\affiliation{Department of Electrical and Computer Engineering, University of Rochester, Rochester, New York, 14627, USA}\affiliation{Department of Physics and Astronomy, University of Rochester, Rochester, New York, 14627, USA}
\author{F\`{e}lix Casanova}
\email[Corresponding author: ]{f.casanova@nanogune.eu}
\affiliation{CIC nanoGUNE, 20018 Donostia-San Sebastian, Basque Country, Spain}
\affiliation{IKERBASQUE, Basque Foundation of Science, 48011 Bilbao, Basque Country, Spain}

\begin{abstract}
\small{Injection of spins into semiconductors is essential for the integration of the spin functionality into conventional electronics. Insulating layers are often inserted between ferromagnetic metals and semiconductors for obtaining an efficient spin injection, and it is therefore crucial to distinguish between signatures of electrical spin injection and impurity-driven effects in the tunnel barrier. Here we demonstrate an impurity-assisted tunneling magnetoresistance effect in nonmagnetic-insulator-nonmagnetic and ferromagnetic-insulator-nonmagnetic tunnel barriers. In both cases, the effect reflects on/off switching of the tunneling current through impurity channels by the external magnetic field. The reported effect is universal for any impurity-assisted tunneling process and provides an alternative interpretation to a widely used technique that employs the same ferromagnetic electrode to inject and detect spin accumulation.}
\end{abstract}

\pacs{}
\maketitle

For the realization of semiconductor spintronic devices \cite{Kato_Science04, Nagaosa_03, Crooker_Science05, Zutic_RMP04, Dery_Nature07, Zutic_RMP04, Appelbaum_Nature07, Dery_Nature07,Li_PRL13, Sasaki_APL11}, the conductivity mismatch problem \cite{Johnson_PRB87,Schmidt_PRB00,Rashba_PRB00,Fert_PRB01} and the difficulty of manipulating semiconductors at the nanoscale are the main issues delaying the progress of this research field. Employing the so-called three-terminal (3T) setup and making use of a single ferromagnetic/insulator contact for both injection and detection of spin-polarized currents was a big step towards this purpose \cite{Dash_Nature09}. Due to the simplicity of the micron-sized structures employed, this setup has gained popularity in semiconductor spintronics \cite{Dash_Nature09, Tran_PRL09, Li_NatureComm11, Dash_PRB11, Aoki_PRB12, Jain_PRL12, Uemura_APL12, Sharma_PRB14, Shiogai_PRB14,Tinkey_arxiv14}. The Lorentzian-shaped magnetoresistance (MR) effect measured in 3T-semiconductor devices has been often attributed to spin injection on accounts of the resemblance to the celebrated Hanle effect in optical spin injection experiments \cite{Optical_Orientation}. However, it has been increasingly realized that the MR reported depends much on the tunneling process and too little on the semiconductor \cite{Dash_Nature09, Tran_PRL09, Li_NatureComm11, Dash_PRB11, Aoki_PRB12, Jain_PRL12, Uemura_APL12, Sharma_PRB14, Shiogai_PRB14, Tinkey_arxiv14}. Furthermore, the typical junction working conditions employed for these measurements, with bias voltage settings much larger than the Zeeman energy, render the signal detection prone to subtle effects driven by impurities embedded in the tunnel barrier \cite{Song_Dery_FIN, Tran_PRL09}.

In this Letter, we elucidate the physics behind such experiments by focusing on the tunnel barrier. Accordingly, our devices render a compact geometry with an aluminum-oxide tunnel barrier created between metallic electrodes, M$_1$/AlO$_x$/M$_2$, as sketched in Fig.~1(a). The M$_1$/AlO$_x$/M$_2$ devices were fabricated in-situ in a UHV electron-beam evaporation chamber with integrated shadow masks. The base pressure of the chamber is below $10^{-9}$ mbar. The thickness of the top and bottom metallic electrodes, M$_1$ and M$_2$, ranged between 10 nm and 15 nm. To decisively probe the role of impurities in the oxide, a series of devices were fabricated with 1) O$_2$ plasma exposure at $10^{-1}$ mbar at a power ranging from around 24 to 40 W for 120 seconds to 210 seconds to minimize the impurity density, or 2) $n-$step ($n$ from 2 to 5) deposition of a 6 {\AA} Al layer with subsequent oxidation of 20 min at $10^{-1}$ mbar of O$_2$ pressure with no plasma. The latter method allows us to vary the density and locations of impurities \cite{Txoperena_APL13, Rippard_PRL02}. The area of the tunnel barrier ranges from 200$\times$275 $\mu$m$^2$ to 375$\times$555 $\mu$m$^2$. The junction resistance $R=V(0)/I$ is measured with the typical 4-point sensing configuration shown in Fig.~1(a), and the associated MR signal $\delta R(B) \equiv [V(B)-V(0)]/I$ is the ratio between the voltage change across the junction and the constant current between the metallic leads when an external magnetic field $B$ is applied. The total amplitude of $\delta R(B)$ will be called $\Delta R$. By using metallic electrodes, we avoid the complications brought by the Schottky barrier and Fermi-level pinning when using a semiconductor \cite{Sze_Book}, and we are able to establish a direct relation between the measured signals and the tunnel barrier. Moreover, we detect similar MR effects in ferromagnetic-insulator-nonmagnetic (FIN) and nonmagnetic-insulator-nonmagnetic (NIN) devices, and explain both of them by considering the magnetic-field-induced on/off switching of the tunneling current through impurities embedded in the tunnel barrier. This important finding calls for investigation of a novel effect and provides an alternative interpretation to recent 3T spin injection experiments, whose magnetoresistance has been attributed to spin accumulation on a nonmagnetic material. Although we do not rule out spin injection in our FIN devices, spin accumulation is clearly not being measured in our setup, since the measured signals are many orders of magnitude higher than those expected from the standard theory of spin diffusion and accumulation \cite{Txoperena_APL13}.

\begin{figure*}[t]
\includegraphics[width=15.5cm]{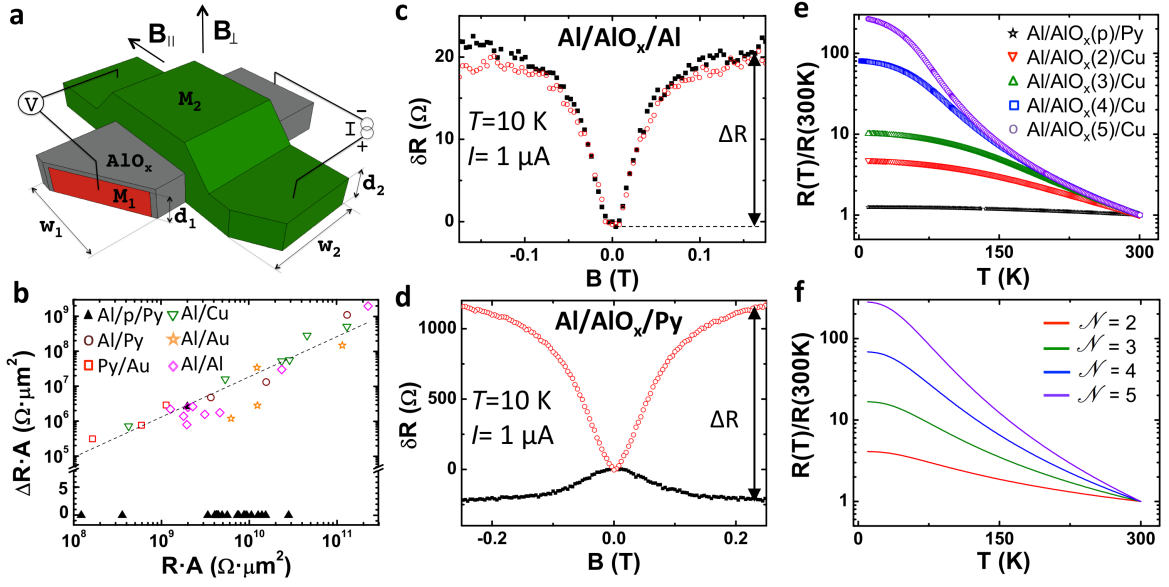}
\caption{Sketch of a tunnel junction, its MR signals and electrical characterization. (\textbf{a}), Scheme of the device and its operation conditions, with the electrode dimensions tagged. (\textbf{b}), $\Delta R \cdot$A as a function of the $R \cdot$A product for different NIN and FIN devices, measured at 10 K and optimum bias conditions for each device. All the tunnel barriers are $n-$step (open symbols), except from the ones labeled as Al/p/Py, which have plasma-oxidized tunnel barriers (solid symbols). Dashed black line is an exponential fit to the data. (\textbf{c}), $\delta R(B)$ of the NIN device for out-of-plane (solid symbols) and in-plane (empty symbols) fields measured at 10~K and 1~$\mu$A, being $R(0)=$13.7~k$\Omega$ under these conditions. (\textbf{d}), $\delta R(B)$ of the FIN device measured at 10~K and 1~$\mu$A (injection from M$_2$$\,$=$\,$Py into M$_1$$\,$=$\,$Al), with $R(0)=$158.9~k$\Omega$. (\textbf{e}), Normalized $R(T)$ for a plasma-oxidized barrier, Al/AlO$_x$(p)/Py, and $n-$step barriers, Al/AlO$_x(n)$/Cu, with $n=$2, 3, 4 and 5. All the data have been measured at 1 $\mu A$. (\textbf{f}), Theoretical $R(T)$ curves due to $\mathcal{N}-1$ phonon-assisted hops through chains of $\mathcal{N}$  impurities. The temperature dependence is governed by the sum of phonon emission ($n_q +1$) and absorption ($n_q$), where $n_q$ is the Bose-Einstein phonon distribution.}\label{fig:fig1}
\end{figure*}

Figure~1(b) shows a compilation of the total amplitude of the MR effect multiplied by the total area of the tunnel barrier ($\Delta R \cdot$A) for $n-$step tunnel barriers (with $n=$2, 3, 4 and 5) with a variety of metallic electrodes, as well as Al/AlO$_x$/Py plasma-oxidized tunnel junctions. For plasma-oxidized AlO$_x$, M$_1$=Al and M$_2$=Py are used (21 devices in total); and the combinations of M$_1$ and M$_2$ metals for $n-$step AlO$_x$ are: M$_1$=Al with M$_2$=Py (3 devices), with M$_2$=Al (9 devices), with M$_2$=Cu (6 devices) and with M$_2$=Au (4 devices), and M$_1$=Py combined with M$_2$=Au (3 devices). Excluding the vast majority of the plasma-oxidized barriers, we find a power law scaling relation between $\Delta R \cdot$A and $R \cdot$A, with an exponent factor of 1.19~($\pm$0.09) [dashed line in Fig. 1(b)]. In the following we focus on the results of two representative impurity-rich NIN (Al/AlO$_x$/Al) and FIN (Al/AlO$_x$/Py) devices whose tunnel barriers are fabricated by a three-step deposition procedure. Figure~1(c) shows $\delta R(B)$ of the NIN device modulated by out-of-plane ($B_\perp$) and in-plane ($B_\|$) fields. The full width at half maximum (FWHM) of both curves is 0.065~T and the junction resistance increases with $B$ regardless of its orientation. We corroborated the isotropy of $\delta R(B)$ in the NIN device for more magnetic field orientations \cite{supple}. Figure~1(d) shows the respective measurements in the FIN device where the FWHM is 0.134~T (0.142~T) and the resistance increases (decreases) when applying an in-plane (out-of-plane) magnetic field. Notably, the FWHM and $\Delta R/R$ values in our devices are comparable to the recurring values seen by 3T-FIN devices employing various insulators and N materials \cite{Dash_Nature09, Tran_PRL09, Li_NatureComm11, Dash_PRB11, Aoki_PRB12, Jain_PRL12, Uemura_APL12, Sharma_PRB14, Erve_NatureNano12, Han_NatureComm13}.

The fact that we observe a non-zero MR signal in NIN, where no spin-polarized source is present, indicates that the MR effect is governed by the oxide barriers rather than by non-equilibrium spin accumulation in N. To better understand the underlying tunnel mechanism, Fig.~1(e) shows the temperature dependence of $R$ in a series of devices with different tunnel barriers. The $R(T)$ of the plasma-oxidized junction shows a weak temperature dependence, in agreement with direct tunneling transport \cite{Akerman_JMMM2002}. In contrast, the data corresponding to $n-$step barriers ($n=$2, 3, 4 and 5) show a stronger T dependence. This dependence can be described by acoustic phonon-assisted tunneling through impurities that dominate the conduction and should follow $R(T) \propto [ \int_0^{\varepsilon_M} d \varepsilon (2n_q(T) +1) \varepsilon^2 ]^{\mathcal{N}-1}$, where $\mathcal{N}$ is the number of impurities assisting the tunneling event, $n_q(T)=1/(e^{\varepsilon / k_B T}-1)$ is the Bose-Einstein distribution, and $\varepsilon_M$ is the upper energy of acoustic phonons in the barrier. Figure~1(f) shows that for an $n-$step tunnel junction we indeed reproduce the experimental results with $\varepsilon_M$$\,$$\sim$$\,$17~meV \cite{Heid_PRB00} and $n=\mathcal{N}$, in agreement with the fabrication method employed. We further support the phonon-assisted tunneling picture by employing the Glazman-Matveev theory \cite{Glazman_JETP88_phonon} to analyze the I-V curves \cite{supple}. Confirmation that the effect is entirely impurity-driven comes from the fact that the MR effect is observed in impurity-rich $n-$step tunnel barriers while being suppressed in plasma-oxidized barriers where direct tunneling is dominant [Fig. 1(b)]. The $T$ and $V$ dependence of the MR amplitude $\Delta R$, displayed in Fig.~2, can be explained in this framework, as will be discussed below. Figures~2(a) and 2(b) show a pronounced decrease of $\Delta R$ with $T$ for the NIN and FIN devices, respectively. Figures~2(c) and 2(d) show that, in both NIN and FIN, $\Delta R$ follows a similar voltage dependence as $R$, except for a sharp decrease when $V$ is close to zero. We observe similar voltage dependences for different $n-$step barriers \cite{supple}.

\begin{figure}[t]
\includegraphics[width=8.5cm]{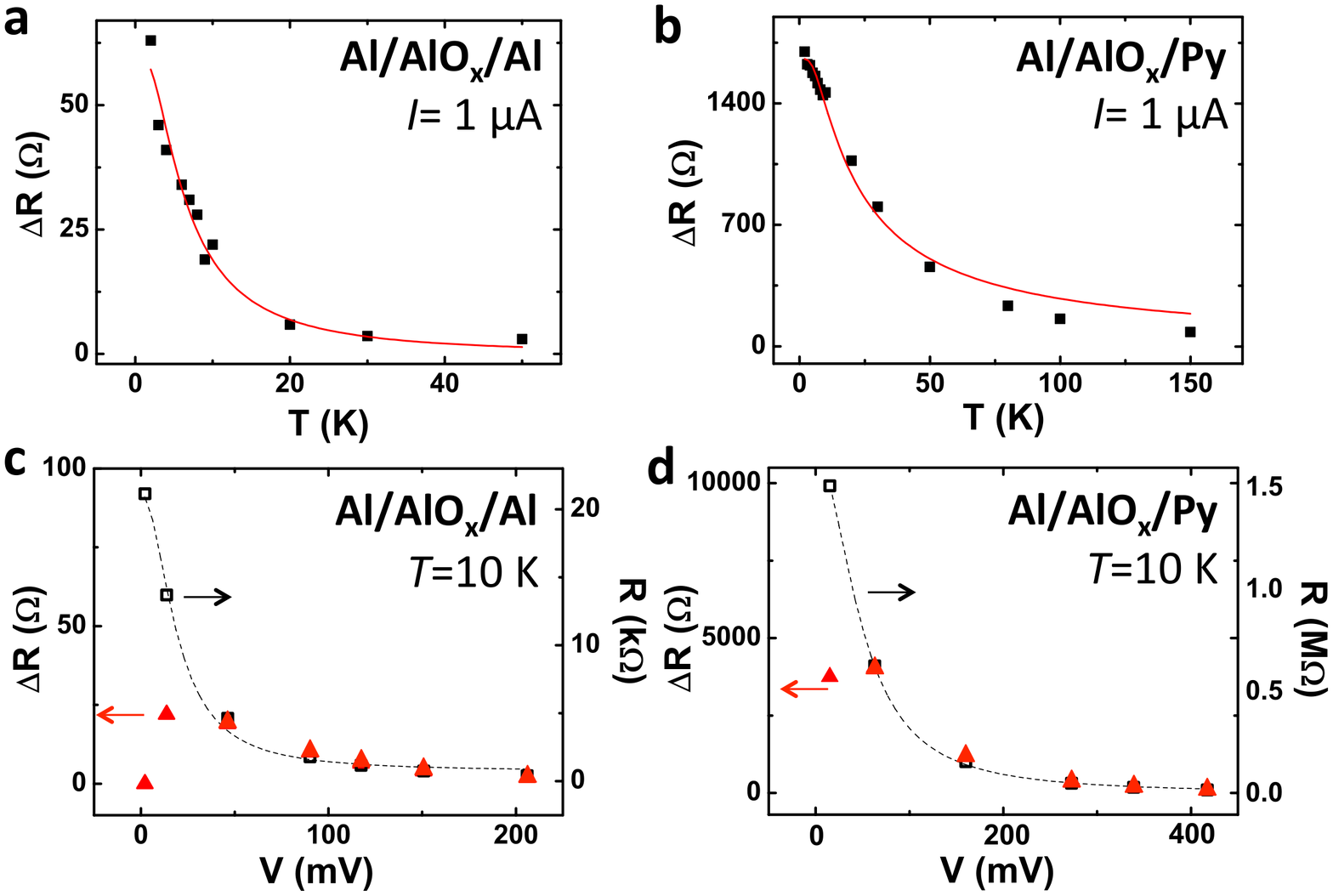}
\caption{Temperature and voltage dependence of the MR amplitude. (\textbf{a}) and (\textbf{b}),
$\Delta R(T)$  measured at 1~$\mu$A for the NIN and FIN devices, respectively. The NIN junction bias voltage changes from 14~mV at 2~K to 8~mV at 50~K, and the FIN one from 160~mV at 2~K to 120~mV at 150~K. Red solid lines are Arrhenius fits to the data with activation energies of $(0.72 \pm 0.07)$ meV for the NIN device and $(1.55 \pm 0.09)$ meV for the FIN device (see text). (\textbf{c}) and (\textbf{d}), The respective values of $\Delta R(V)$ and $R(V)$ measured at 10~K. The signals are symmetric for $V>0$ and $V<0$. The black dashed line is a guide to the eye. } \label{fig:fig2}
\end{figure}

We propose a tunneling mechanism to explain the experimental findings. Using the gained information regarding tunneling across impurity chains in our devices, we classify impurities with large on-site Coulomb repulsion energy ($U \gg eV$)  into type A and type B classes. In type A (B), the filling energy for the first (second) electron is within the bias window \cite{Glazman_JETP88_Coulomb,Bahlouli_PRB94}. This simple classification of the energetic levels of the localized states captures the core physics of our experiments \cite{Boero_PRL97}. Figure~3(a) shows an example when both types form an A-B chain in the tunnel barrier of a NIN junction. When electrons tunnel in the direction from A to B, this chain enables on (off) current switching in small (large) external magnetic fields. To understand this effect, we first focus on the steady-state spin configuration in the chain. Once an electron tunnels from the left bank into the type A impurity, it can be intuitively viewed as an ideal polarized source (`one electron version of a half metal'). Due to Pauli blocking, this electron cannot hop to the second level of the type B impurity if the first level of the latter is filled with an electron of same spin orientation [see Fig. 3(a)]. The steady-state current across the chain is therefore blocked. This blockade can be lifted when the correlated spin configuration is randomized by spin interactions, which include the spin-orbit coupling \cite{Prioli_PRB95}, hyperfine coupling with the nuclear spin system \cite{Boero_PRL97}, and spin-spin exchange interactions with unpaired electrons in neighboring impurities \cite{Helman_PRL76}.  Whatever is the dominant interaction, we can invoke a mean-field approximation and view this interaction as an internal magnetic field at the impurity site that competes with the external field. When the external field is much larger than the internal fields, the type A and type B impurities in the chain see similar fields and the current is Pauli blocked as explained before. In the opposite extreme of negligible external field, the blockade is lifted since the correlated spin configuration is violated by spin precession about internal fields that are likely to point in different directions on the A and B sites. This behavior is illustrated by Fig.~3(b). Although A-B impurity chain is the simplest case that supports magnetic field modulation of the current, similar modulations will also occur in longer chains containing an A-B sequence.

\begin{figure*}[t]
\includegraphics[width=15cm]{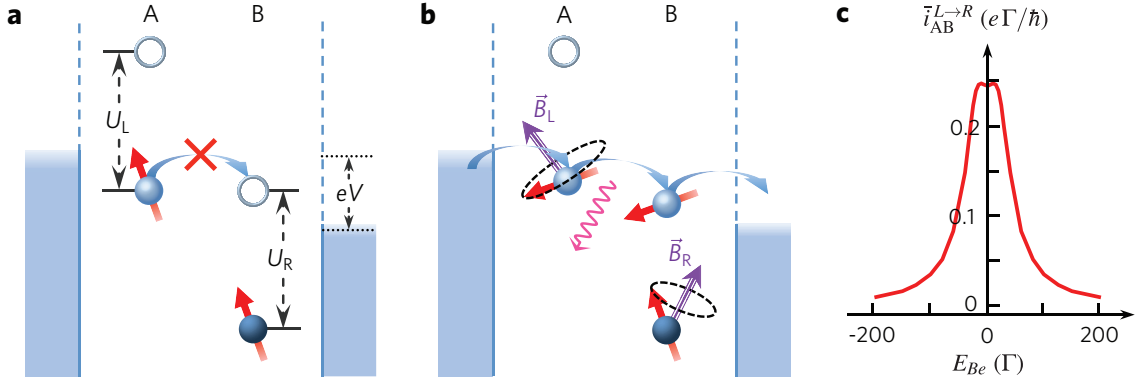}
\caption{Schematics for impurity-assisted MR mechanisms and the theoretical result. (\textbf{a}), A-B impurity chain in the bias window of a NIN junction. Due to the large on-site Coulomb repulsion ($U_{\ell} \gg eV$), the current across the chain is Pauli blocked when the electron spins of the lower levels in A and B are parallel. (\textbf{b}), The current blocking is lifted when different magnetic fields in A and B randomize the correlated spin orientation of the chain (see text). The dominant tunneling process between two impurities is assisted by phonon emission. All the rest of the possible two-impurity chains (B-A, A-A, B-B) do not modulate the current in the NIN junctions \cite{supple}. The A-B impurity chain analyzed in this figure also modulates the current in FIN devices. (\textbf{c}), Theoretical calculation of the current across A-B impurity chain as a function of external magnetic field for a NIN device [see text after Eq.~(\ref{eq:dR})].} \label{fig:fig3}
\end{figure*}

Next we consider FIN junctions. Due to the magnetization of F, there are two main differences compared to NIN. First, the polarized tunnel current in FIN facilitates partial blocking of the impurity-assisted current already without an external field. In NIN junctions, on the other hand, the current is unblocked without an external field due to the randomized spin configuration induced by the presence of internal fields. As will be explained below, the result is that in FIN junctions the tunnel resistance can either increase (larger blocking) or decrease (weaker blocking) depending on the magnetic field orientation with respect to the magnetization axis of F.  The second difference is that chains with at least one A-B sequence are needed in order to have field modulation in NIN (where the type A impurity plays the role of `polarizing' the incoming current). In case of FIN, on the other hand, a single impurity is sufficient to block the current. It can be any chain with at least one type B impurity when electrons flow from F to N (spin injection), or at least one type A impurity when electrons flow from N to F (spin extraction) \cite{Song_Dery_FIN}.  Current blockade is established once the spin in the lower level of the type B (A) impurity is parallel (antiparallel) to the majority spins of F in spin injection (extraction). The blockade is lifted when applying an out-of-plane field whose magnitude is much smaller than the saturation field of F. Spin precession of the electron in the lower level of the type B (A) impurity lifts the blockade since this electron can no longer keep a parallel (antiparallel) spin configuration with the majority spins of F. This physical picture explains the measured reduction in the resistance of the FIN for this field orientation [see Fig. 1(d)]. On the other hand, by applying a field parallel to the magnetization axis of F, the resistance increases since the external field impedes spin precession induced by random internal magnetic fields. Therefore, the current blocked configurations are reinforced: spins in the lower levels of type B (A) impurities are parallel (antiparallel) to the majority spins of F in injection (extraction). Such reinforcement is equivalent to the behavior of NIN junctions under a magnetic field pointing in any direction. The above discussed behavior in FIN explains the measured anisotropy in $\delta R(B)$ shown in Fig.~1(d). Finally, we emphasize that, details aside, the underlying physics of the MR effect is the same in both FIN and NIN junctions.

To quantify the impurity-assisted tunneling magnetoresistance effect, we describe a toy model based on the tunneling through two-impurity chains by generalizing the Anderson impurity Hamiltonian model to our tunneling case \cite{Glazman_JETP88_phonon}. The steady-state current across the impurity chains are then found by invoking non-equilibrium Green function techniques and deriving master equations in the slave-boson representation \cite{Zou_PRB88,LeGuillou_PRB95}. The technical details are given in the supplemental material \cite{supple}. The steady-state current essentially represents competition between the Zeeman terms, impurity-lead coupling ($\Gamma_{\ell}$ where $\ell$ denotes Left/Right impurity-lead pair), and inter-impurity coupling ($\Gamma_{dd}$). These coupling terms reflect tunneling rates (via $\hbar/\Gamma$). Solving the master equations for the particular case of the A-B impurity chain and bias setting described in Fig.~3, we obtain the following steady-state solution for the dominant contribution \cite{supple},
\begin{eqnarray}\label{eq:current_NIN}
i^{\scriptscriptstyle L\rightarrow R}_{\scriptscriptstyle AB} (\theta)
\approx
\frac{2q}{\hbar} \left(
\frac{1}{\Gamma_L }+ \frac{1}{\Gamma_R }
-\frac{1}{ \Gamma_L+\Gamma_R}
 +\frac{4}{\Gamma_{dd} \sin^2\theta}\right)^{-1}.
\end{eqnarray}
This expression describes the magnetic-field modulated current via an A-B impurity chain, where the magnetic field dependence is manifested via the angle $\theta=\theta_R-\theta_L$. For large enough external field ($\mathbf{B}_e$) the effective fields in the left and right impurities are aligned ($\mathbf{B}_{L} \!\parallel \!\mathbf{B}_{R}$), and the current is blocked (i.e., $\theta\rightarrow 0$ leading to $i^{\scriptscriptstyle L\rightarrow R}_{\scriptscriptstyle AB} \rightarrow 0$). When $B_e$ is much smaller than the internal fields, on the other hand, $\langle \sin^2\theta \rangle $ is effectively of the order of 1/2 after averaging over the distribution of $\theta$, and the current can flow. The full expression for $i^{\scriptscriptstyle L\rightarrow R}_{\scriptscriptstyle AB}$ is given in Eq.~(S3) of the supplemental material \cite{supple}, and in Eq.~(\ref{eq:current_NIN}) above we show its simplified form in the limit that the Zeeman energy is larger than the impurity-lead and impurity-impurity couplings ($\Gamma$'s). This limit is generally satisfied due to the random distribution of internal  fields whose magnitudes and variations can readily exceed those of the weak coupling parameters. In this limit, the FWHM are determined by the characteristic amplitude of the internal fields. This explains why the stray fields due to the F/I roughness \cite{Dash_PRB11} that add to the internal fields in FIN give rise to somewhat larger FWHM values compared to NIN. It also justifies the independence of the measured FWHM values on the thickness of the tunnel barrier. Equation~(\ref{eq:current_NIN}) shows a series-like resistance for the A-B chain where the negative term, $-1/(\Gamma_L+\Gamma_R)$,  stems from the coherence between two impurities \cite{supple}.

We can now recover the measured signal by noting that
\begin{eqnarray} \label{eq:dR}
\frac{\delta R(\mathbf{B}_e)}{R} =  N_{\text{AB}}\times \frac{\bar{i}^{\,\,L \rightarrow R}_{\text{AB}}}{I},
\end{eqnarray}
where $ N_{\text{AB}}$ is the number of A-B chains with $U_{\ell} \gg eV$, and $I$ is the total current enabled via tunneling over impurity clusters with various sizes and on-site repulsion $U$'s.

All the obtained experimental results are readily understood by applying the above analysis. First, Fig.~3(c) shows a current simulation using Eq.~(\ref{eq:current_NIN}) after averaging over the amplitude and orientation of the internal fields. Since the tunneling probability decays exponentially with the barrier thickness, the dominant contribution comes from equidistant impurities for which $\Gamma_L = \Gamma_R = \Gamma_{dd} = \Gamma$ \cite{Bahlouli_PRB94}. Using this equality, we model the internal field in each of the impurities as an independent normalized Gaussian distribution whose mean and standard deviation are 20$\Gamma$ and $6\Gamma$, respectively \cite{supple}. We observe that the shape of the simulated curve is in agreement with the Lorentzian shape measured in both NIN and FIN [Figs.~1(c) and 1(d)]. Second, we explain the $\Delta R(T)$ behavior for the NIN and FIN devices. On the one hand, we observe a stronger T dependence of the signal for NIN than for FIN [see Figs.~2(a) and 2(b)]. The origin for this behavior is that in NIN devices the blockade is effective when $U_{\ell} \gg eV$ for both impurities on the A-B chain. By contrast, in the FIN devices, it is sufficient to have one such impurity due to the spin polarization of F, rendering $\Delta R$ less temperature dependent. Using this information, $\Delta R(T)$ can be fitted by a typical Arrhenius law $\delta R(T) \propto \left[1 - \exp{(-E_a / k_BT)}\right]^{m}$ where $m=2(1)$ for NIN (FIN) devices. The red lines in Figs.~2(a) and 2(b) show the dependence where the activation energy is $E_a = 0.72 \pm 0.07$~meV for the NIN device and $E_a = 1.55 \pm 0.09$~meV for the FIN device. The activation energy $E_a\!\sim\! 1$ meV is associated with the threshold of small impurities to merge into larger clusters resulting in $U \lesssim eV$ \cite{Helman_PRL76}. This scenario is compatible with our devices where apart from isolated impurities, we might also have impurities in close proximity behaving as big clusters as temperature is increased. Third, the decrease of $\Delta R(V)$ at low bias values [Figs. 2(c) and 2(d)] is because of the vanishing number of A-B channels within the small bias window. Finally, related to that, the relative signal $\Delta R/R$ is a result of the small portion of A-B chains with $U_{\ell} \gg eV$ among all cluster chains. The fact that $\Delta R/R$ is nearly constant comparing all devices, as shown in Fig.~1(b), is in agreement with Eq.~(\ref{eq:dR}).

In conclusion, the MR effect shows how the impurity-assisted tunnel resistance can be modulated by a magnetic field when the Zeeman splitting of the impurity spin states is smaller compared to the applied bias voltage. Other impurity-driven effects reported up to date, such as the Kondo effect or Coulomb correlation in resonant tunneling \cite{Gordon_Nature98,Ephron_PRL92, Xu_PRB95}, appear in the opposite regime at strong magnetic fields. This mechanism therefore promises new possibilities to explore local states in disordered materials or nanostructures. Our analysis puts NIN and FIN junctions on an equal footing, with the physical picture readily generalizable to chains with $\mathcal{N}\geq2(1)$ impurities in NIN~(FIN) junctions. This novel magnetoresistance effect is general for any impurity-assisted tunneling process regardless of the oxide thickness or materials used. Therefore, the presented work will be used as a benchmark to spin injection experiments to any nonmagnetic material, and specially will redirect research of semiconductor spintronics, with all the implications in such a technologically relevant area.

The authors acknowledge Dr. A. Bedoya-Pinto for fruitful discussions. The work in Spain is supported by the European Union 7th Framework Programme (NMP3-SL-2011-263104-HINTS, PIRG06-GA-2009-256470 and the European Research Council Grant 257654-SPINTROS), by the Spanish Ministry of Economy under Project No. MAT2012-37638 and by the Basque Government under Project No. PI2011-1.  The work in USA is supported by NRI-NSF, NSF, and DTRA Contracts No. DMR-1124601, ECCS-1231570, and HDTRA1-13-1-0013, respectively.

\begin{widetext}
\setcounter{figure}{0}
\renewcommand{\thefigure}{S\arabic{figure}}
\setcounter{equation}{0}
\renewcommand{\theequation}{S\arabic{equation}}
\renewcommand{\thebibliography}{S\arabic{bibliography}}
\linespread{1.5}

\setlength{\parindent}{1cm}
\setlength{\parskip}{0.4cm}
\linespread{1.5}
\begin{center}
\Large{Supplemental Material for ``Universal impurity-assisted tunneling}
\Large{magnetoresistance under weak magnetic field''}
\\*
\vspace{0.7cm}
\normalsize{Oihana Txoperena, Yang Song, Lan Qing, Marco Gobbi,}
\\*
\normalsize{Luis E. Hueso, Hanan Dery, F\`{e}lix Casanova}

\end{center}
\large{

\begin{figure}[b]
\includegraphics[width=0.4\textwidth]{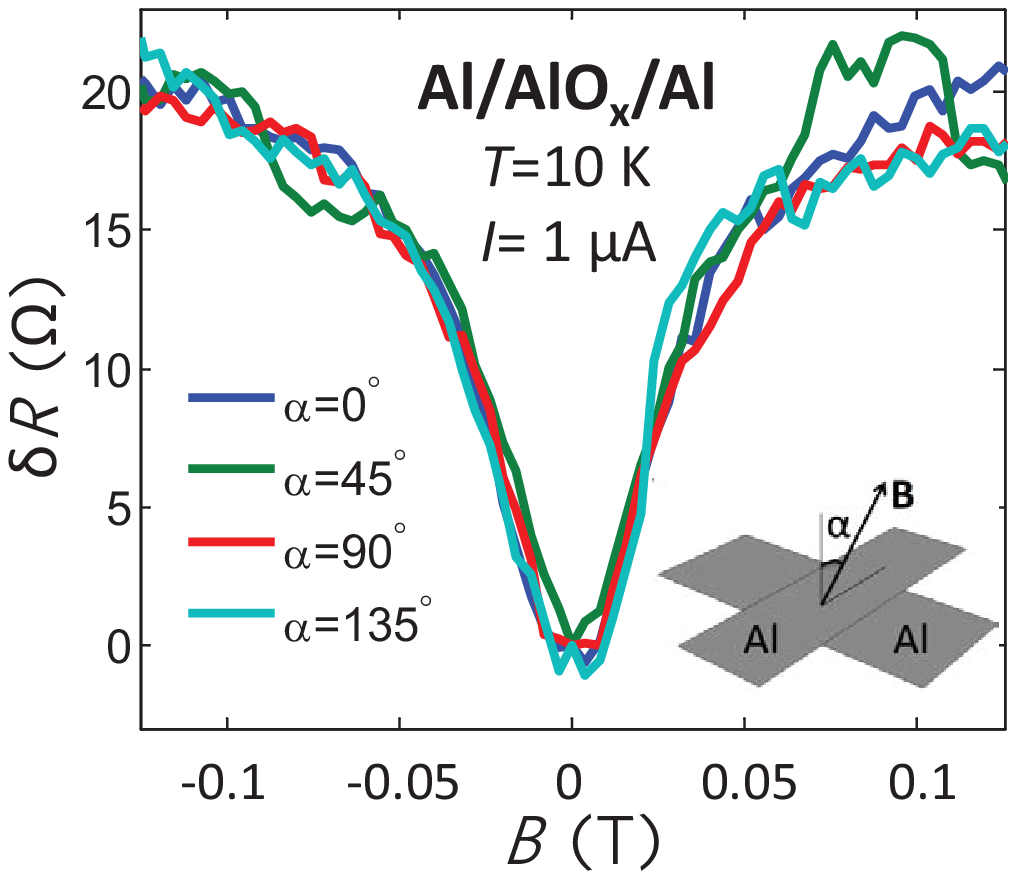}
\caption{ The isotropy of MR signal in NIN device with respect to magnetic field directions. Four different magnetic field directions (see inset) are applied on the  Al/AlO$_x$(3-step)/Al device, which we focus on in the main text. The measurements are done at 10 K and 1 $\mu$A. Due to Pauli blocking across A-B chains, the tunnel resistance increases with the field regardless of its orientation (see main text).
 } \label{fig:figS1}
\end{figure}

\noindent \textbf{Additional experimental results and discussion}

In Fig. S1, we show two additional magnetic field (B) directions applied on the representative NIN device. We observe no correlation between the MR signals and the $\mathbf{B}$ field directions.

\begin{figure}[b]
\includegraphics[width=0.6\textwidth]{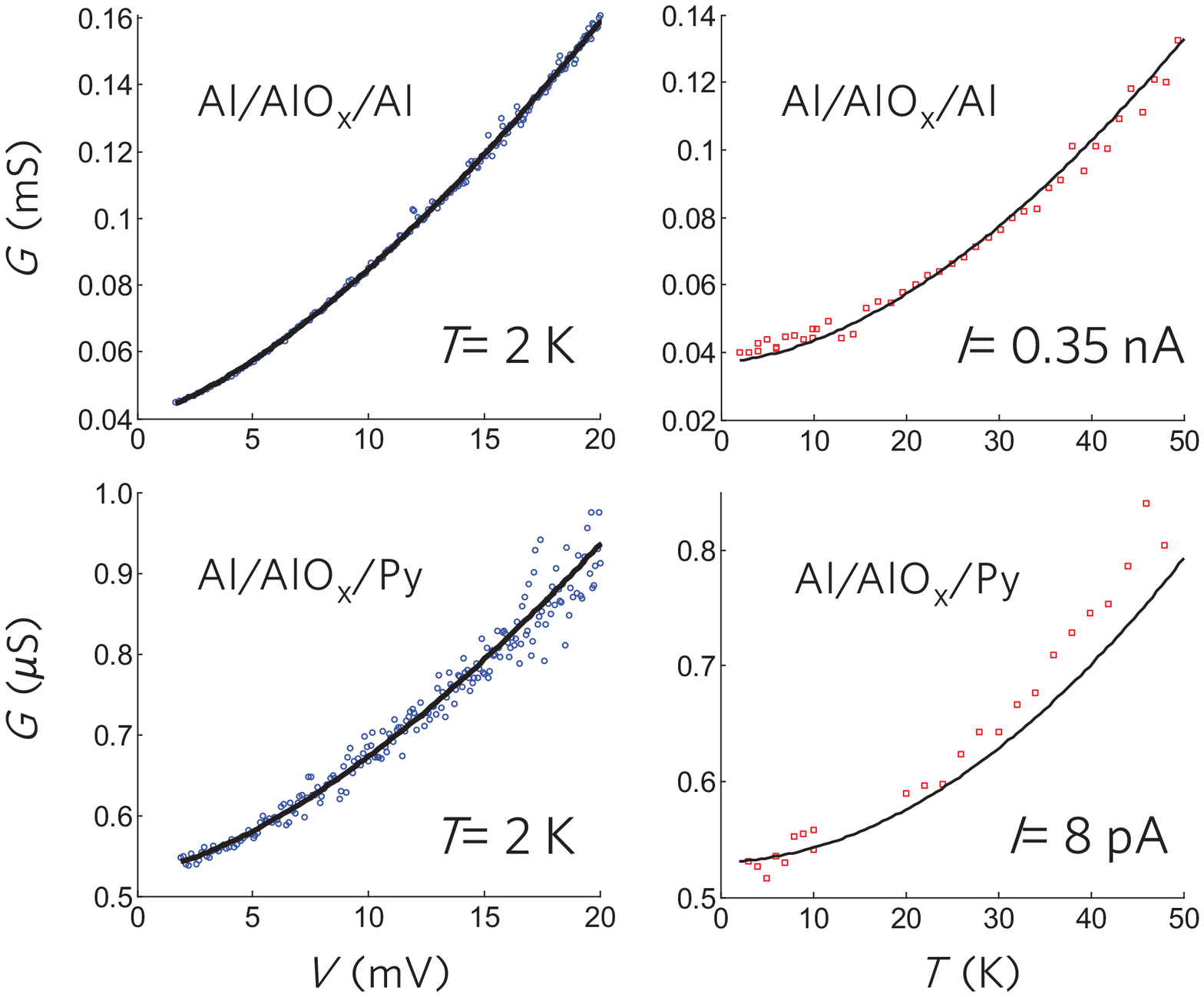}
\caption{Total conductance $G=1/R$ for the representative 3-step NIN and FIN devices as a function of voltage and temperature under small bias windows. The voltage dependence is measured at $eV\gg k_B T$, and the temperature dependence is measured at $k_B T\gg eV$ where $V$ is on the order of 10 $\mu$V in both devices. Circles are measured data and solid lines are the theoretical fitting for phonon-assisted tunneling via impurities (see text).} \label{fig:figS2}
\end{figure}

Figure S2 characterizes the conductance of the tunnel junctions, $G$, for the representative 3-step NIN and FIN devices at small bias windows. The left panels show their voltage dependence at $eV\gg k_B T$ while the right panels show their temperature dependence at $k_B T\gg eV$. In these regimes we can apply Glazman-Matveev theory for ordinary hopping via impurity chains [S1,S2]. We fit the obtained data by $G(V)=c_1+c_2V^p$ and $G(T)=c_3+c_4 T^p$, where $p=\mathcal{N}-2/(\mathcal{N}+1)$ with $\mathcal{N}$ being the average impurity number in the chains under these small bias windows. From the voltage-dependent measurements, we obtain $\mathcal{N}=2.088\pm 0.008$ for the NIN sample and $\mathcal{N}=2.12\pm 0.04$ for the FIN sample. From the temperature-dependent ones, we obtain $\mathcal{N}=2.39\pm0.06$ for the NIN sample and $\mathcal{N}=2.2\pm0.2$ for the FIN sample. Therefore, the results obtained from the voltage- and temperature-dependent measurements are consistent, and show that at these small bias windows the transport in our 3-step tunnel barriers is dominated by conduction through two-impurity chains, meaning $\mathcal{N} \approx n-1$, where $n$ is the number of steps of the tunnel barrier. Note that in such small bias window condition, we can observe a perceivable background of $V$ and $T$ independent conductance due to direct and resonant tunneling, which becomes negligible in the usual working condition (e.g. 1 $\mu$A constant current in Fig. 1(c) and 1(d) of the main text). The average impurity number $\mathcal{N}$ slowly increases as the bias window increases [S2], as under the condition used in Fig. 1(e) of the main text where the number of impurity is closer to  the number of deposition steps, $\mathcal{N} \approx n$. Finally, we point out that the above theoretical results [S1] are applicable only when max$\{eV,k_BT\}\lesssim \varepsilon_M$, where $\varepsilon_M$ is the maximum acoustic phonon energy. We obtain that $\varepsilon_M$ is on the order of 17 meV from Fig. 1(e) and 1(f) of the main text. Note that in Fig. 1(e) we have $eV\!>\!\{k_B T, \varepsilon_M\}$, and in this case the only important temperature dependence comes from that of the phonon population. It is also worth mentioning that the existence of impurities in the tunnel barrier can be also manifested as resonant peaks in the second derivative of the I(V) curve [S3]. Tinkey \textit{et al.} have recently performed inelastic electron tunneling spectroscopy measurements (IETS) in SiO$_2$ tunnel barriers and observe some sharp peaks not corresponding to Si or SiO$_2$ vibrational modes in the IETS spectra. They observe that the peaks disappear as the barrier thickness is increased, which results in an increase of the barrier resistance. This happens due to the increase in the density of impurities when increasing the barrier thickness, resulting in a dense energetic distribution of impurities and disappearance of the peaks. After doing the corresponding calculations and comparison of our tunnel barrier resistance-area products with those in Ref.~[S3], we conclude that the impurity-density in our tunnel barriers is too high for such peaks to be observed in the IETS.

\begin{figure}
\includegraphics[width=\textwidth]{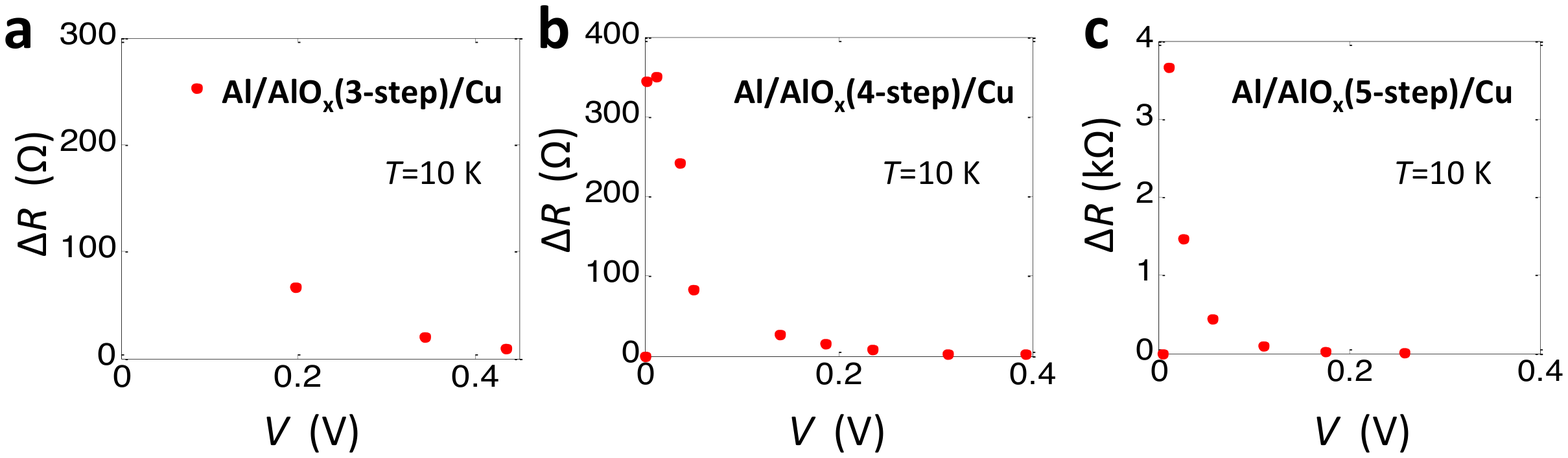}
\caption{ Bias dependence of the MR signal $\Delta R$ for Al/AlO$_x$/Cu devices, with (\textbf{a}) 3-step, (\textbf{b}) 4-step and (\textbf{c}) 5-step tunnel barriers. All the measurements are done at 10 K.
} \label{fig:figS3}
\end{figure}

Figure~S3 shows $\Delta R(V)$ of several Al/AlO$_x(n)$/Cu samples, with comparable voltage dependences to those shown in Fig.~2 of the main text. At large voltage values, $\Delta R$ follows the same voltage dependence as the total tunnel resistance, as explained by Eq.~(2) of the main text. However, when the bias window reduces significantly, the small portion of impurity chains that are subject to magnetic field modulation (i.e., chains with an A-B sequence) becomes basically non-available compared to the total impurity chains, as well as to the direct and resonant tunneling channels. As a result, there are sharp drops of $\Delta R$ as the voltage is close to zero. It is worth pointing out that the voltage value where $\Delta R$ is maximum, $V_{max}$, decreases as $n$ increases, with $V_{max}$=85 mV, 12 mV and 5 mV for $n=$3, 4 and 5, respectively. This can be qualitatively explained as follows: the higher the $n$ is, the longer the impurity chains are, and the more probable is to find A-B chains fulfilling $U_{\ell} \gg eV$, obtaining MR at smaller bias values.

\clearpage
\noindent \textbf{Theory and numerical calculations}

We describe the tunneling through two-impurity chains by generalizing the Anderson impurity Hamiltonian model to our tunneling case [S1],
\begin{eqnarray}
\!H \!\!&\!=\!&\!\!
\sum_{\ell \sigma} \left[(E_{d\ell} + \sigma E_{B_\ell} \cos\theta_\ell)
n_{d\ell \sigma} +
E_{B_\ell} \sin\theta_\ell d^\dag_{\ell\sigma} d_{\ell\bar\sigma}\right]
+ \sum_\ell U_\ell n_{d\ell \uparrow}  n_{d\ell \downarrow}
\nonumber\\
&&
\!\!
+ \sum_{\ell\mathbf{k}\sigma} \!\left[\varepsilon_{\mathbf{k}\ell\sigma}
n_{\mathbf{k}\ell\sigma}
\!+\!(  V_{d\mathbf{k}\ell} \mathbf{k}^\dag_{\ell\sigma} d_{\ell'\sigma}\!\!
+\!\textrm{h.c.}  )\right]
+ \!\sum_{\mathbf{q}}\!\!\left[
\varepsilon_{\mathbf{q}} n_\mathbf{q}
\!+\! V_{dd} (\lambda_\mathbf{q} \mathbf{q}^\dag\! +\!\textrm{h.c.})
\sum_{\sigma} (d^\dag_{L\sigma}d_{R\sigma} \!+\!\textrm{h.c.} ) \right].
\end{eqnarray}
$\sigma\!=\!-\bar\sigma\!=\! \pm1$ denotes spin and $\ell\!=\!\{L,R\}$ are for Left/Right leads or impurities. $d$ ($\mathbf{k}$)  denotes impurity (lead) electrons and $\mathbf{q}$ denotes phonons.  The respective energy levels and occupation operators are $\{E_{d,\ell}$, $\varepsilon_{\mathbf{k}\ell\sigma}, \varepsilon_{\mathbf{q}}\}$,  and  $\{n_{d\ell\sigma} \equiv d^\dag_{\ell\sigma} d_{\ell\sigma},\,n_{\mathbf{k}\ell\sigma} \equiv \mathbf{k}^\dag_{\ell\sigma} \mathbf{k}_{\ell\sigma},\,n_{\mathbf{q}} \equiv \mathbf{q}^{\dag}\mathbf{q}\}$. $U_{\ell}$ is the on-site Coulomb repulsion energy. The Zeeman splitting energy at the $\ell$-th impurity is $2E_{B_\ell} \equiv g\mu_B B_\ell$, where $\mathbf{B}_{\ell}$ is the sum of internal and external magnetic fields. $\mathbf{B}_L$ and $\mathbf{B}_R$ define the $xz$ plane and  $\theta_\ell$ is the angle between $\mathbf{B}_{\ell}$ and the $z$ axis.  $V_{dd}$ and $V_{d\mathbf{k}\ell}$ give rise to coupling between two impurities and with their nearby leads, and  $\lambda_\mathbf{q}$ is the electron-phonon interaction matrix element. We have kept only linear inelastic tunneling terms which dominate the resonant tunneling at our finite bias [S1].

\vspace{0.2cm}
\noindent \underline{Master equations and the full analytical expression}

To find the steady-state current across the impurity chains from the above Hamiltonian, we invoke non-equilibrium Green function techniques and derive master equations in the slave-boson representation [S4,S5]. They describe the competition between the Zeeman terms, impurity-lead coupling ($\Gamma_{\ell}$) and inter-impurity coupling ($\Gamma_{dd}$). The latter two in the weak coupling regime are expressed by $\Gamma_{\ell} \!=\!2\pi \!\sum_\mathbf{k} |V_{\ell \mathbf{k}}|^2 \delta(E_{d\ell}\!-\!\varepsilon_{\mathbf{k}\ell})$ and $\Gamma_{dd }\!=\! 4\pi \!\sum_\mathbf{q}|V_{dd}\lambda_q|^2\delta(\Delta E_{d}\!-\!\varepsilon_{\mathbf{q}})$, respectively, where $\Delta E_d\!=\!E_{dL}\! -\! (E_{dR}+U_R)$.  Below we derive the general master equations for the A-B tunneling chain, under arbitrary magnetic fields at the two impurity sites and phonon population $n_{\mathbf{q}}$. We focus on the dominant contribution for which $\Delta E_d\gg k_B T$ (i.e. $n_q\ll 1$).

The basis states can be understood as  $|n_{L}  n_{R}\rangle$, where $n=\{0,\uparrow,\downarrow, 2\}$ has four possible states. We define density operators by $\hat\rho^{n_{L}  n_{R}}_{m_{L}  m_{R}}\equiv |n_{L}  n_{R}\rangle \langle m_{L}  m_{R}|$ and $\hat\rho_{n_{L}  n_{R}}\equiv \hat\rho^{n_{L}  n_{R}}_{n_{L}  n_{R}}$.
Without loss of generality, we set $\theta_L=0$,  $\theta_R=\theta_R-\theta_L=\theta$, and choose $E_{dL}>E_{dR}$.
\begin{subequations}\label{eq:master_B}
\begin{eqnarray}
\hbar\frac{d }{d t}\rho_{0\sigma}
&=&
-2\Gamma_L \rho_{0\sigma} + \Gamma_R \rho_{02} -2E_{B_R}\sin\theta {\rm Im}\rho^{0\bar\sigma}_{0\sigma},
\\
\hbar\frac{d }{d t}\rho_{02}
&=&
-2(\Gamma_L +\Gamma_R)\rho_{02}
\nonumber\\
&&+ \Gamma_d[-2n_\mathbf{q} \rho_{02}+ (n_\mathbf{q}+1)(\rho_{\uparrow\downarrow}+\rho_{\downarrow\uparrow}-2{\rm Re}\rho_{\uparrow\downarrow}^{\downarrow\uparrow})] ,\quad
\\
\hbar\frac{d }{d t}\rho_{\sigma\bar\sigma}
&=&
\Gamma_L \rho_{0\bar\sigma} +\Gamma_R \rho_{\sigma2}
+2E_{B_R}\sin\theta {\rm Im}\rho^{\sigma\bar\sigma}_{\sigma\sigma}
\nonumber\\
&&+ \Gamma_d[n_\mathbf{q} \rho_{02}+ (n_\mathbf{q}+1)(-\rho_{\sigma\bar\sigma}+{\rm Re}\rho_{\sigma\bar\sigma}^{\bar\sigma \sigma})],\quad
\\
\hbar\frac{d }{d t}\rho_{\sigma2}
&=&
 \Gamma_L \rho_{02} -2\Gamma_R \rho_{\sigma2},
\quad
\\
\hbar\frac{d }{d t}\rho_{\sigma\sigma}
&=&
\Gamma_L \rho_{0\sigma} + \Gamma_R \rho_{\sigma2}
- E_{B_R}\sin\theta {\rm Im}\rho^{\sigma\bar\sigma}_{\sigma\sigma},
\quad
\\
\hbar\frac{d }{d t}\rho^{\downarrow\uparrow}_{\uparrow\downarrow}
&=&\!\!
\Gamma_d [-n_\mathbf{q} \rho_{02}+ (n_\mathbf{q}+1) (\rho_{\uparrow\downarrow}+\rho_{\downarrow\uparrow}-2\rho_{\uparrow\downarrow}^{\downarrow\uparrow})/2]
\nonumber\\
&&\!\!\!\!+2i (\!-\!E_{B_L}\!+\!E_{B_R}\cos\theta)\rho^{\downarrow\uparrow}_{\uparrow\downarrow}
+i E_{B_R}\sin\theta(\rho^{\downarrow\downarrow}_{\uparrow\downarrow}\!-\!\rho^{\downarrow\uparrow}_{\uparrow\uparrow}),
\quad
\\
\hbar\frac{d }{d t}\rho^{0\downarrow}_{0\uparrow}
&=&
 -2\Gamma_L \rho^{0\downarrow}_{0\uparrow} + i E_{B_R}[-2\cos\theta \rho^{0\downarrow}_{0\uparrow} +\sin\theta (\rho_{0\uparrow}-\rho_{0\downarrow}) ],\qquad
\\
\hbar\frac{d }{d t}\rho^{\downarrow2}_{\uparrow2}
&=&
-2\Gamma_R \rho^{\downarrow2}_{\uparrow2} -2 i E_{B_L}\rho^{\downarrow2}_{\uparrow2},
\qquad
\\
\hbar\frac{d }{d t}\rho^{\downarrow\downarrow}_{\uparrow\uparrow}
&=&\!\!
-2i (E_{B_L}\!+\!E_{B_R}\cos\theta) \rho^{\downarrow\downarrow}_{\uparrow\uparrow}
+i E_{B_R} \sin\theta (\rho^{\downarrow\uparrow}_{\uparrow\uparrow}-\rho^{\downarrow\downarrow}_{\uparrow\downarrow}),
\quad
\\
\hbar\frac{d }{d t}\rho^{\sigma\bar\sigma}_{\sigma\sigma}
&=&
\Gamma_L \rho^{0\bar\sigma}_{0\sigma} + \Gamma_d (n_\mathbf{q}+1)(\rho^{\bar\sigma\sigma}_{\sigma\sigma} -\rho^{\sigma\bar\sigma}_{\sigma\sigma})/2
  \nonumber\\
&&+ i E_{B_R}[-2\sigma\cos\theta \rho^{\sigma\bar\sigma}_{\sigma\sigma} +\sin\theta (\rho_{\sigma\sigma}-\rho_{\sigma\bar\sigma}) ],
\quad
\\
\hbar\frac{d }{d t}\rho^{\bar\sigma\sigma}_{\sigma\sigma}
&=&
\Gamma_L \rho^{\bar\sigma2}_{\sigma2}+ \Gamma_d (n_\mathbf{q}+1)(\rho^{\sigma\bar\sigma}_{\sigma\sigma} -\rho^{\bar\sigma\sigma}_{\sigma\sigma})/2
   \nonumber\\
&&-2\sigma i E_{B_L}\rho^{\bar\sigma\sigma}_{\sigma\sigma}
+i E_{B_R}\sin\theta (\rho^{\bar\sigma\bar\sigma}_{\sigma\sigma}-\rho^{\bar\sigma\sigma}_{\sigma\bar\sigma}) ],\quad
\end{eqnarray}
\end{subequations}
The equations are not all independent but supplemented by $1=\rho_{02}+\sum_\sigma (\rho_{0\sigma}+ \rho_{\sigma\bar\sigma} +\rho_{\sigma2} +\rho_{\sigma\sigma})$. Having solutions to all matrix elements at $n_q\ll 1$, and $I = \frac{q}{\hbar} 2 \Gamma_L (\rho_{0\uparrow}+\rho_{0\downarrow}+\rho_{02})$, we get
\begin{eqnarray}\label{eq:current_full}
I =\frac{8q}{\hbar} (\Gamma_L+\Gamma_R)\Gamma_d\Gamma_L\Gamma_R
E_{B_L}^2E_{B_R}^2(E_{B_L}^2-E_{B_R}^2)^2   \sin^2\theta/\Lambda,
\end{eqnarray}
where
\begin{eqnarray}\label{eq:deno}
\Lambda
\!&\!=\!&\!
(E_{B_L}^2\!+\!E_{B_R}^2\!+\!2E_{B_L}E_{B_R}\cos\theta)\Gamma^2_d\Gamma_L\Gamma_R(\Gamma_L\!+\!\Gamma_R)
\times
\nonumber\\
\!&&\!\big[E_{B_L}^4\!-\!E_{B_L}^2\!E_{B_R}^2(1\!+\!\cos^2\!\theta) \!+\!
E_{B_R}^4\big]
\!+\! 4E_{B_L}^2\!E_{B_R}^2 \!(E_{B_L}^2 \!-\! E_{B_R}^2)^2
\nonumber\\
&&\times\big[\Gamma_d(\Gamma_L^2\!+\!\Gamma_L\Gamma_R\!+\!\Gamma^2_R)\sin^2\theta
+4\Gamma_L\Gamma_R(\Gamma_L\!+\!\Gamma_R)\big],\nonumber
\quad
\end{eqnarray}
Equation~(S3) leads to the approximated form in Eq.~(1) of the main text. The negative term in Eq.~(1) is a consequence of physical invariance under the rotation of spin coordinate, and it also occurs in B-A, A-A and B-B chains whose currents are independent of magnetic field. It is reflected in the off-diagonal elements in the master Eqs.~(S2).

\vspace{0.2cm}
\noindent\underline{Calculation of averaged current expressions via AB chains, as well as on BA, AA, and BB chains}

In the following we show how to obtain the averaged current expression plotted in Fig.~3(c) starting from the full current expression via an AB chain shown in Eq.~(S2). To do that, we need to integrate over the internal field distributions at the two impurities, taking into account that they experience local internal magnetic fields due to spin interactions in addition to the external field. In order to do the integration, we express $E_{B_L}$, $E_{B_R}$ and $\sin\theta$ in Eq.~(S2) in terms of the left and right internal fields $\mathbf{B}_{i_L}$ and $\mathbf{B}_{i_R}$, and external field $\mathbf{B}_{e}$. If $z$ direction is set along $\mathbf{B}_{e}$, from $\mathbf{B}_{L(R)} = \mathbf{B}_{i_{L(R)}}+ \mathbf{B}_{e}$ one can obtain
\begin{eqnarray}
B_{\ell}= \sqrt{B^2_{i_{\ell}}+B^2_e+2B_{i_\ell} B_e \cos\theta_{i_\ell}},
\quad
\cos\theta_{\ell} =  \frac{B_{i_\ell}\cos\theta_{i_\ell}+B_e}{B_\ell}, \quad
{\rm and}
\quad \phi_\ell = \phi_{i_\ell},
\end{eqnarray}
where $\ell=L,R$.  The angle $\theta$ between $\mathbf{B}_{L}$ and $\mathbf{B}_{R}$ can be expressed as follows
\begin{eqnarray}
\cos\theta= \cos\theta_L\cos\theta_R + \sin\theta_L\sin\theta_R\cos(\phi_L- \phi_R ).
\end{eqnarray}
As previously mentioned, the averaged current is a result of integration over internal field distribution probability $\mathcal{F}_\ell(B_{i_\ell}, \theta_\ell,\phi_\ell)$,
\begin{eqnarray}\label{eq:i_AB_integration}
\bar{i}_{AB} =
\int d^3\mathbf{B}_{i_L} \int d^3\mathbf{B}_{i_R}\left(\mathcal{F}_L \times \mathcal{F}_R \times i_{AB}\right),
\end{eqnarray}
where $\int d^3\mathbf{B}_\ell \mathcal{F}_\ell= 1$ and, for simplicity, we assume that $\mathcal{F}_L$ and $\mathcal{F}_R$ are independent. For example, we may assume they are Gaussian distributions with finite variation around a mean value on the radial direction. Figure 3(c) is obtained in this way by a straightforward numerical integration of Eq.~(S6). We assume $\Gamma_L=\Gamma_R=\Gamma_{dd}$ because at this condition the impurity-assisted inelastic tunneling current is maximum [S1,S2].

For the purpose of gaining more insight of the magnitude of the signal and its trend with external magnetic field, we can make justified simplifications in order to carry out analytical integration. Since we are interested mainly in the regime of average internal field and its variation much larger than the tunneling rate, $\{E_{B_L}, \! E_{B_R},\! |E_{B_L}\!-\!E_{B_R}|\}\gg\{\Gamma_R,\!\Gamma_L,\!\Gamma_{dd}\}$, we can properly use the approximation in Eq.~(1) of the main text. Doing so, for any $\mathcal{F}_{L,R}$ with spherical symmetry, at $B_e=0$ we have
\begin{eqnarray}
\bar{i}_{AB}(B_e=0) &\approx& \frac{e}{\hbar}
\frac{8\pi^2\Gamma_d \Lambda_1}
{\Lambda_2}
\left[1-\frac{\Lambda_1{\rm arctanh}(\sqrt{\frac{\Lambda_2}{\Lambda_2+\Lambda_1}}) }
{\sqrt{\Lambda_2(\Lambda_2+\Lambda_1)} }  \right]
\int^\infty_0  d B_{i_L} B_{i_L}^2 \mathcal{F}_L(B_{i_L})\int^\infty_0  d B_{i_R} B_{i_R}^2 \mathcal{F}_R(B_{i_R})
\nonumber\\
&=&\frac{e}{\hbar}
\frac{\Gamma_d \Lambda_1}
{2\Lambda_2}
\left[1-\frac{\Lambda_1{\rm arctanh}(\sqrt{\frac{\Lambda_2}{\Lambda_2+\Lambda_1}}) }
{\sqrt{\Lambda_2(\Lambda_2+\Lambda_1)} }  \right]
\end{eqnarray}
where
\begin{eqnarray}
\Lambda_1=2\Gamma_L \Gamma_R(\Gamma_L +\Gamma_R), \quad
\Lambda_2=\frac{1}{2}\Gamma_d(\Gamma_L^2+\Gamma_L \Gamma_R +\Gamma_R^2).
\end{eqnarray}
We have $\bar{i}_{AB}(B_e=0)\approx 0.257 \Gamma e/\hbar$ well matching the numerical result in Fig. 3(c), with the corresponding parameters used $\Gamma_L=\Gamma_R=\Gamma_{dd}=\Gamma$ and $\mathcal{F}_{\ell}\propto \exp[-(E_{Bi_\ell}-20\Gamma)^2/2(6\Gamma)^2]$.

In order to obtain an approximate but analytical trend of the current as a function of external field $B_e$, we can further approximate by using $i_{AB}\approx \Gamma_d \sin^2\theta e/2\hbar$, obtaining
\begin{eqnarray}
\bar{i}_{AB}(B_e) &\approx& \frac{e}{2\hbar}\Gamma_d
\int d^3\mathbf{B}_{i_L} \int d^3\mathbf{B}_{i_R}\mathcal{F}_L(B_{iL}) \mathcal{F}_R (B_{iR})
 \nonumber\\
 && \left\{1-\left[ \frac{B_{i_L}\cos\theta_{i_L}+B_e}{B_L}   \frac{B_{i_R}\cos\theta_{i_R}+B_e}{B_R} + \frac{B_{i_L}\sin\theta_{i_L}}{B_L}\frac{B_{i_R}\sin\theta_{i_R}}{B_R}\cos(\phi_{iL}- \phi_{iR} )\right]^2\right\}
\nonumber\\
&=&  \frac{e}{2\hbar}\Gamma_d -  \pi^2\frac{e}{\hbar}\Gamma_d \bigg[\int^\infty_0\!\!\!\! d {B}_{i_L}\! \!\!\int^\infty_0\!\!\!\! d {B}_{i_R} \mathcal{F}_L(B_{iL}) \mathcal{F}_R (B_{iR})({B}_{i_L}{B}_{i_R})^2
\nonumber\\
&&
\qquad\qquad\qquad\qquad\int^1_{-1}\!\!\!d z_L \!\!\int^1_{-1}\!\! d z_R
\frac{2 (B_{i_L}z_L+B_e)^2 (B_{i_R}z_R+B_e)^2+ B^2_{i_L}(1-z_L^2) B^2_{i_R}(1-z_R^2)}
{(B^2_{i_{L}}+B^2_e+2B_{i_L} B_e z_L) (B^2_{i_R}+B^2_e+2B_{i_R} B_e z_R)}\bigg]
\nonumber\\
&=&  \frac{e}{2\hbar}\Gamma_d \bigg\{ 1-  \frac{\pi^2}{8}
 \left(76-\frac{100}{3}\right)  \left(\int^\infty_0\!\!\!\! d {B}_{i}\mathcal{F}(B_{i}){B}_{i}^2\right)^2
 \nonumber\\
 &&\qquad-\frac{\pi^2}{8}\frac{1}{B_e^6}\left(\int^\infty_0\!\!\!\! d {B}_{i}\mathcal{F}(B_{i})
  {B}_{i}
\left[\frac{2}{\sqrt{3}} B_e B_{i}(5B_e^2-3B_{i}^2)-\sqrt{3}(B_e^2-B_{i}^2)^2 \ln \frac{|B_e-B_{i}|}{B_e+B_{i}}
\right] \right)^2
\bigg\}
\nonumber\\
&\approx&  \frac{e}{\hbar}\Gamma_d \bigg\{ \frac{1}{3}
-\frac{1}{768B_e^6\bar{B}^2_i}  \left(2B_e \bar{B}_i(3\bar{B}_i^2-5B_e^2)+3(\bar{B}_i^2-B_e^2)^2  \ln \frac{|B_e-\bar{B}_{i}|}{B_e+\bar{B}}\right)^2
\bigg\}.
\end{eqnarray}
where in the last step we have used the condition that the mean magnetic field $\bar{B}_i$ of the distribution $\mathcal{F}_{i\ell}$ is much larger than its standard deviation, and replaced $\ln [|B_e-B_{i}|/(B_e+B_{i})]$ by $\ln [|B_e-\bar{B}_{i}|/(B_e+\bar{B}_{i})]$ in the integrand (this excellent approximation has been checked numerically for the whole range of $B_e/\bar{B}_i$).

Last, we show explicitly that the current via other two-impurity chain types BA, AA and BB is magnetic field independent for the NIN devices. They are obtained by exactly solving similar master equations as those shown Eq.~(S2).
\begin{eqnarray}
i_{BA} &=& \dfrac{e}{\hbar}
\dfrac{2\Gamma_d\Gamma_L\Gamma_R(\Gamma_L+\Gamma_R) (n_\mathbf{q}+1) }{\Gamma_L\Gamma_R(\Gamma_L+\Gamma_R+2\Gamma_d n_\mathbf{q})+2\Gamma_d(\Gamma_L^2+\Gamma_L\Gamma_R+\Gamma^2_R)(n_\mathbf{q}+1)},
\\
i_{AA} &=&\dfrac{e}{\hbar}\dfrac{ 2 \Gamma_d \Gamma_L\Gamma_R (2\Gamma_L+\Gamma_R)(n_\mathbf{q}+1)}
{2\Gamma_L\Gamma_R(2\Gamma_L+\Gamma_R + 2\Gamma_d n_\mathbf{q})+\Gamma_d(4\Gamma_L^2+2\Gamma_L\Gamma_R+\Gamma_R^2)(n_\mathbf{q}+1) },
\\
i_{BB}
&=& \dfrac{e}{\hbar}\dfrac{2\Gamma_d \Gamma_L\Gamma_R(\Gamma_L+2\Gamma_R)  (n_\mathbf{q}+1)}
{ 2\Gamma_L\Gamma_R(\Gamma_L+2\Gamma_R + 2\Gamma_d n_\mathbf{q}) +\Gamma_d(\Gamma_L^2+2\Gamma_L\Gamma_R+4\Gamma_R^2)(n_\mathbf{q}+1)}.
\end{eqnarray}

\noindent \textbf{Discussion of new experimental results from other Groups}

The mechanism proposed in the main text can readily explain the recurring magnetic field modulated signals from most 3T-FIN experiments when N is a semiconductor.  The commonly seen feature that the MR signals depend less on the lead materials but more on the barriers is consistent with our framework.  Here we manifest the versatility of this theory by explaining in detail some new observations that cannot be explained by previous theories [S6].

First, the Fe/MgO/Si devices in Ref.~[S6] show coexistence of two effects: the total tunnel resistance $R$ saturates when the oxide barrier becomes ultrathin, whereas the MR signal $\Delta R$ shows a robust exponential dependence on the thickness of the oxide barrier. This phenomena can be straightforwardly explained by our tunneling theory of a FIN junction via one impurity, which is most likely to reside at the impurity-prone atomic interface between the oxide (MgO) and the Schottky barrier in the Si region.  The total tunnel resistance saturation is a typical behavior of resonant tunneling via a single impurity. When the thickness $t$ of the MgO barrier is significantly reduced ($\ln \Gamma_{\rm Fe}\!\propto\! -t$), the resonant resistance is dominated by the smaller tunneling rate $\Gamma_{\rm Si}\,\,(\ll \Gamma_{\rm Fe})$  yielding
\begin{eqnarray}
I_{\rm res}    =   \frac{2e}{\hbar}\frac{\Gamma_{\rm Fe} \Gamma_{\rm Si}}{\Gamma_{\rm Fe} + \Gamma_{\rm Si}}
\approx
\frac{2e}{\hbar}\Gamma_{\rm Si}.
\end{eqnarray}
Therefore, the tunneling resistance becomes constant, since $\Gamma_{\rm Si}$ is determined by the Schottky barrier which remains the same regardless of the MgO thickness.

Concerning the magnitude of the MR current $\Delta i$, we can see that when $\Gamma_{\rm Si}\ll \Gamma_{\rm Fe}$ it depends on the relative difference of the spin-dependent resonant tunneling rates, $p^2\Gamma_{\rm Si}/(1-p^2)\Gamma_{\rm Fe}$, in addition to  $\Gamma_{\rm Si}$. $\Gamma_{\rm Fe}=(\Gamma_{\rm Fe\uparrow}+\Gamma_{\rm Fe\downarrow})/2$ and $p$ are the spin-average tunneling rate and the spin polarization of the ferromagnet, and the $p^2$ dependence stems from the additional current-induced spin polarization of the impurity. Since ferromagnets are not pure half metal and $p$ is a fraction of 1, the intuitive current on/off picture induced by the magnetic field becomes effectively the difference of resonant tunneling rates from two opposite spins.  When the tunnel barrier is really thin, their difference  becomes extremely small because even the minority  spin resonant tunneling rate between Fe and the impurity is much larger than that between Si and the impurity. As a result, the bottleneck of the `current-blocked state' is caused by the Schottky barrier in the Si region rather than the minority spin tunneling from Fe. So the magnetic field modulation tends to be suppressed in this limit. More quantitative analysis is found in  Ref.~[S7].

All in all, we have
\begin{eqnarray}
\Delta i \propto \Gamma_{\rm Si}^2/\Gamma_{\rm Fe}.
\end{eqnarray}
Combining $\Delta i$ and $I_{\rm res}$ with Eq.~(2) of the main text, we have $\Delta R\propto 1/\Gamma_{\rm Fe}$ for $\Gamma_{\rm Si}\ll \Gamma_{\rm Fe}$.  In the other limit of thick MgO barrier ($\Gamma_{\rm Si}\gg \Gamma_{\rm Fe}$), both $I_{\rm res}$ and $\Delta i$ are proportional to $\Gamma_{\rm Fe}$.  The magnetic modulation in this case is effective where $\Delta i$ is governed by the large relative difference $p^2$.   Again, we have $\Delta R\propto 1/\Gamma_F$. The two proportionality prefactors are of the same order of magnitude. Our MR picture leads to the overall exponential dependence of $\Delta R$ on the oxide thickness. This dependence has also been seen by other groups [S8]. The remaining control experimental results of Ref.~[S7] are naturally understood within this MR picture. Replacing Si with metal eliminates the impurity-prone MgO/Si interface and Schottky barrier,  therefore suppressing the impurity-assisted TMR. Inserting non-magnetic layer into the Fe/MgO interface turns off the spin polarization source [S8], and again suppresses our one-impurity-assisted TMR effect in FIN devices.

A noteworthy difference between FIN junctions where N is a semiconductor or a metal is the absence of the Schottky barrier in the latter ones. In both cases, treating the tunnel barrier by plasma oxidation is likely to eliminate the impurities inside the oxide layer. But defects can still be produced on the atomic interface between two different materials. When N is a semiconductor, the resonance current via impurities on the atomic interface between the oxide and the Schottky barrier can be large, and the MR signal comes from those impurities with large on-site Coulomb repulsion. When N is a metal, on the other hand, the tunnel current via impurities at the oxide/metal interface is negligible because the density of states in N is much higher than that of the impurities. This difference can then explain the negligible MR signal in our plasma-oxidized samples where both leads are metallic, and its appearance in experiments when N is a semiconductor.
}

\vspace{0.5cm}

\normalsize{

[S1] L. I. Glazman and  K. A. Matveev, Zh. Eksp. Teor. Fiz.  \textbf{94}, 332 (1988) [Sov. Phys. JETP \textbf{67}, 1276 (1988)].

[S2] Y. Xu, D. Ephron, M. R. Beasley, Phys. Rev. B \textbf{52}, 2843 (1995).

[S3] H. N. Tinkey, P. Li, I. Appelbaum, arXiv: 1405.2297 (2014).

[S4] Z. Zou and P. W. Anderson, Phys. Rev. B \textbf{37}, 627(R) (1988).

[S5] J. C. Le Guillou and E. Ragoucy, Phys. Rev. B \textbf{52}, 2403 (1995).

[S6] S. Sharma \textit{et al.}, Phys. Rev. B \textbf{89}, 075301 (2014).

[S7] Y. Song and H. Dery, Phys. Rev. Lett. \textbf{113}, 047205 (2014).

[S8]	T. Uemura, K. Kondo, J. Fujisawa, K.-I. Matsuda, M. Yamamoto, Appl. Phys. Lett. \textbf{101}, 132411 (2012).
}

\end{widetext}

\end{document}